\documentstyle[aps,pra,epsfig,twocolumn]{revtex}
\newcommand{\be}{\begin{equation}}
\newcommand{\ee}{\end{equation}}
\newcommand{\bea}{\begin{eqnarray}}
\newcommand{\eea}{\end{eqnarray}}
\def\C{{\cal C}}

\newcommand{\ket}[1]{\mbox{$| #1 \rangle$}}

\begin{document}
\draft

\title{Interaction cost of non-local gates}

\author{G. Vidal$^{1}$, K. Hammerer$^{1}$ and J. I. Cirac$^{2}$}

\address{
$^1$Institut f\"ur Theoretische Physik, Universit\"at Innsbruck,
A-6020 Innsbruck, Austria\\
$^2$ Max--Planck Institut f\"ur Quantenoptik, Hans--Kopfermann
Str. 1, D-85748 Garching, Germany
}

\date{\today}

\maketitle

\begin{abstract}

We introduce the interaction cost of a non-local gate as the minimal time of interaction required to perform the gate when assisting the process with fast local unitaries. This cost, of interest both in the areas of quantum control and quantum information, depends on the specific interaction, and allows to compare in an operationally meaningful manner any two non-local gates. In the case of a two-qubit system, an analytical expression for the interaction cost of any unitary operation given any coupling Hamiltonian is obtained.
One gate may be more time-consuming than another for any possible interaction. This defines a partial order structure in the set of non-local gates, that compares their degree of non-locality. We analytically characterize this partial order in a region of the set of two-qubit gates. 

\end{abstract}


An elementary concern in quantum information theory is to stablish the trade-off between different physical resources that are relevant for information processing. A controlled Hamiltonian interaction between quantum systems is one instance of useful resource. It can be employed, for example, to simulate the dynamics of another multipartite quantum system.
On the other hand multi-particle unitary gates are a requirement for universal quantum computation. In particular, two-qubit gates ---together with one-qubit gates--- can be taken as the building block of quantum computers. 

A detailed study of the connections existing between non-local Hamiltonians and non-local gates is thus of interest from a quantum information perspective, but this issue is also relevant in other areas. For instance, the synthesis of multipartite gates from Hamiltonian interactions ---and, in particular, time--minimizing schemes--- has been recently  analyzed in the context of quantum control theory \cite{Khaneja}.
Whereas the requirements for arbitrary manipulation of single qubits are presently met in a number of experimental schemes, the engineering of two-qubit gates can be only (partially) achieved with very few systems \cite{Fort}. In real experiments not only an interaction Hamiltonian between the qubits, but also considerable command on them in order to process the interaction, are required. For instance, mechanisms to switch on and off the interaction, as well as to accurately drive the systems towards the desired joint evolution, are needed. But even from a theoretical perspective, a description of two-qubit gates in terms of interactions able to prescribe optimal protocols for gate synthesis was so far missing. Here we shall provide such a description.

More generally, we consider a set of subsystems with a given Hamiltonian $H$, and assume that arbitrarily fast local unitaries (LU) can be performed to properly tailor the evolution that $H$ induces. The aim is to perform some joint unitary transformation $U$ on the systems. This is the setting considered in \cite{Khaneja} and corresponds to the so-called {\em gate simulation under LU} of \cite{Bennett}. Two definitions are needed to specify the problems that we shall address.

\vspace{2mm}

{\bf Definition 1:} The interaction cost $\C_H(U)$ of a non-local gate $U$ given a Hamiltonian $H$ denotes the minimal time needed in order to perform $U$ using the interaction $H$ and fast LU.

\vspace{2mm}

{\bf Definition 2:} We say gate $U$ is more non-local than gate $V$, and write $V \leq U$, when for all interactions $H$ the interaction cost of U is never smaller than that of $V$,
\be
V \leq U ~~\equiv ~~\C_H(V)\leq \C_H(U) ~~~\forall H.
\ee

\vspace{2mm}

First we shall show how the interaction cost $\C_H$ can be explicitly computed for any gate and any interaction of a two-qubit system. This is possible by considering results recently developed in the areas of quantum control \cite{Khaneja} and quantum information \cite{Bennett,Duer,Kraus,Hammerer}. In \cite{Khaneja} considerable progress towards the solution was made, and only a final optimization was left unsolved. The results of \cite{Bennett,Duer,Kraus,Hammerer} provide the tools needed to perform such an optimization and thereby complete the results of \cite{Khaneja}.

Definition 2 introduces a partial order structure in the set of non-local gates. This structure captures the intuition, in terms of the resources needed to perform a gate, that one gate may be ``more non-local'' than another. Our second result is an analytical characterization of this structure in a region of the set of two-qubit gates.

We start by describing known facts concerning the simulation of non-local Hamiltonians and the synthesis of non-local gates. 

\vspace{2mm}

($i$) {\em Optimal simulation of two-qubit Hamiltonians under LU.} Any Hamiltonian acting on two qubits is uniquely represented, for the purposes of
simulation under LU, by its canonical form \cite{Duer,Bennett}
\be
H = \sum_i h_i \sigma_i\otimes\sigma_i,~~~h_1\geq h_2 \geq |h_3|,
\label{Hami}
\ee
where $\sigma_i$, $i=1,2,3$, stand for the Pauli matrices. 
In the rest of the paper $H$ denotes a Hamiltonian written in its canonical form and $\vec{h}$ denotes the vector $(h_1,h_2,h_3)$ with its properly ordered coefficients. The special majorization relation $\vec{x} \prec_s \vec{y}$ between three dimensional real vectors $\vec{x}$ and $\vec{y}$ is relevant in this context. It is given by the set of inequalities 
\bea
x_1 &\leq& y_1, \nonumber\\
x_1 + x_2 - x_3 &\leq& y_1+y_2-y_3,\nonumber\\
x_1 + x_2 + x_3 &\leq& y_1+y_2+y_3.
\label{smaj}
\eea
where the components $x_i$ and $y_i$ are assumed to fulfill $x_1 \geq x_2 \geq |x_3|$ and $y_1 \geq y_2 \geq |y_3|$ \cite{smaj}.

\vspace{2mm}

{\bf Fact 1 (Theorem of \cite{Bennett}):} The minimal time overhead $t_{H'|H}$ (i.e., the inverse of efficiency $s_{H'|H}$) required to simulate Hamiltonian $H'$ by Hamiltonian $H$ and fast LU is the minimal value of $c\geq 0$ such that the vectors $\vec{h}'$ and $\vec{h}$ satisfy $\vec{h}'\prec_s c\vec{h}$. Protocols for optimal simulation are known.

\vspace{2mm}

($ii$) {\em Optimal synthesis of two-qubit gates under LU.} Any two-qubit gate $U$ can be written in terms of local unitaries $u_i$ and $v_i$ and Pauli matrices $\sigma_k$ as \cite{Khaneja,Kraus}
\be
U = (u_1\otimes v_1) e^{-i \sum_k \lambda_k\sigma_k\otimes\sigma_k} (u_2\otimes v_2).
\label{deco}
\ee
In \cite{Kraus} it is shown how to obtain this decomposition. Notice that gate $U$ is equivalent, up to local unitaries performed on the qubits before and after $U$, to 
\be
U_{\vec{\lambda}} \equiv e^{-i\sum_k \lambda_k \sigma_k\otimes\sigma_k}.
\label{lambdas}
\ee
Since we assume the ability to perform instantaneous (i.e., sufficiently fast) LU operations, the synthesis of $U$ is as time-consuming as that of $U_{\vec{\lambda}}$, and we need only focus on the later. In addition \cite{Hammerer}, to each $U$ there corresponds a unique $U_{\vec{\lambda}^0}$ with $\vec{\lambda}^0\equiv (\lambda_1^0,\lambda_2^0,\lambda_3^0)$ such that $\lambda_1^0\geq \lambda_2^0 \geq |\lambda_3^0|$,~~ $\lambda^0_1,\lambda_2^0 \in [0,\pi/4]$ and $\lambda_3^0 \in (-\pi/4,\pi/4]$, that we shall call its canonical form. In what follows we will often represent any two-qubit gate $U$ by its canonical form $U_{\vec{\lambda}^0}$ or by its corresponding (unique) vector $\vec{\lambda}^0=(\lambda_1^0,\lambda_2^0,\lambda_3^0)$. Recall that all commutators $[\sigma_j\otimes\sigma_j,\sigma_k\otimes\sigma_k]$ vanish, and that $\exp(-i\pi/2 \sigma_j\otimes\sigma_j)=-i\sigma_j\otimes\sigma_j$ is a local gate. This implies that for any vector $\vec{n}=(n_1,n_2,n_3)$ with integer components $n_j$, 
\be
U_{\vec{\lambda}^0} =_{LU} U_{\vec{\lambda}^0}U_{\frac{\pi}{2}\vec{n}}=U_{\vec{\lambda}^0+\frac{\pi}{2}\vec{n}},
\label{alldeco}
\ee
with $\vec{\lambda}^0+\pi/2\vec{n}$ essentially exhausting all vectors compatible with the gate $U_{\vec{\lambda}^0}$ \cite{upto}. [$=_{LU}$ is used to denote equivalence under LU.]. 

In Theorem 10 of \cite{Khaneja} the problem of time-optimally producing a two-qubit gate $U$ using interaction $H$ is shown to reduce to a specific minimization over all possible decompositions of $U$ of the form (\ref{deco}). Here we rephrase the theorem in terms of the notion of Hamiltonian simulation and the concepts introduced before. Without loss of generality, we refer only to unitary operations that can be written as in (\ref{lambdas}), and associate a self-adjoint operator $H_{\vec{\lambda}}\equiv \sum_i \lambda_i \sigma_i \otimes \sigma_i$ to each possible decomposition.

\vspace{2mm}

{\bf Fact 2 (Theorem 10 of \cite{Khaneja}, readapted):} The time-optimal way to synthesize gate $U$ with interaction $H$ and fast $LU$ consists of simulating, among all Hamiltonians $H_{\vec{\lambda}}$ such that $U=\exp(-iH_{\vec{\lambda}})$, the one with smallest time overhead $t_{H_{\vec{\lambda}}|H}$. The minimal interaction time (i.e., the interaction cost $\C_H(U)$) is given by the smallest time overhead $t_{H_{\vec{\lambda}}|H}$.

\vspace{2mm}

Our first aim is to perform the optimization described in Fact 2. This is feasible because we have an analytical characterization both of all (infinitely many) decompositions of $U$ (cf. Eq. (\ref{alldeco})) and of the time overhead $s_{H_{\vec{\lambda}}|H}$ for any decomposition (cf. Fact 1), as expressed in the following lemma.

\vspace{2mm}

{\bf Lemma:}
The interaction cost $\C_H(U)$ is the minimal value $c \geq 0$ such that a vector $\vec{n}$ of integers exists satisfying
\be
\vec{\lambda}^0 + \frac{\pi}{2}\vec{n}\prec_s c \vec{h}.
\ee

\vspace{2mm}

It is useful to introduce, for each $\vec{n}$, the {\em pre-cost} $c_{\vec{n}}$ as the minimal value $c\geq 0$ such that $\vec{\lambda}^0 + \pi/2\vec{n} \prec_s c\vec{h}$. Pre-cost $c_{\vec{n}}$ is the overhead needed to simulate $H_{\vec{\lambda}^0+\pi/2\vec{n}}$ by $H$ or, equivalently, the minimal time $t$ needed to travel, in the set of non-local gates, from the identity operator to $U$ along the path defined by $\vec{\lambda}^0+\pi/2\vec{n}$. Intuitively, a large $\vec{n}$ corresponds to a ``long'' ---and therefore non-optimal--- path. Following this intuition we arrive at our first result.

\vspace{2mm}

{\bf Theorem 1:} The interaction cost $\C_H(U)$ or minimal time needed to create gate $U$ by using Hamiltonian $H$ and fast LU is given by
\be
\C_H(U) = \min\{ c_{(0,0,0)}, ~ c_{(-1,0,0)}\},
\label{theorem1}
\ee
that is, the minimal of two pre-costs, one corresponding to the canonical vector $\vec{\lambda}^0=(\lambda_1^0,\lambda_2^0,\lambda_3^0)$ of $U$ and the other to the vector $(\frac{\pi}{2}- \lambda_1^0,\lambda_2^0,-\lambda_3^0)$ \cite{clar}. The time-optimal protocol consists in simulating the corresponding Hamiltonian (either $\vec{h}_1 \equiv (\lambda_1^0,\lambda_2^0,\lambda_3^0)$ or $\vec{h}_2=(\frac{\pi}{2}- \lambda_1^0,\lambda_2^0,-\lambda_3^0)$) by $H$ for time $t=\C_H(U)$.

\vspace{2mm}

{\em Remark.} Thus, in order to time-optimally perform gate $U$ with Hamiltonian $H$, we can proceed as follows. Using Ref. \cite{Kraus}, we compute $\vec{\lambda}^0$ from $U$, and using Refs. \cite{Duer,Bennett} we compute $\vec{h}$ from $H$. Theorem 1 gives the minimal time of simulation and the Hamiltonian (either $\vec{h}_1$ or $\vec{h}_2$) to be simulated, and finally Ref. \cite{Bennett} describes an optimal protocol for simulating the convenient Hamiltonian by $H$ and LU.

\vspace{2mm}

{\em Proof:} We need to see that $\C_H(U)$ as given by Eq. (\ref{theorem1}) is the minimal pre-cost, i.e. $\C_H(U)\leq c_{\vec{n}}$ for all $\vec{n}$. It is straightforward to check from Eq. (\ref{smaj}) that (T1.$i$) for any two vectors $\vec{x}$ and $\vec{y}$, with components $x_1\geq x_2\geq |x_3|$, $y_1\geq y_2\geq |y_3|$, the minimal $c\geq 0$ such that $\vec{x}\prec_s c\vec{y}$ satisfies $c \leq 3 x_1/y_1$; (T1.$ii$) if $\vec{x} \prec_s \vec{x}'$, then $\vec{x}'\prec_s \vec{y} \Rightarrow \vec{x}\prec_s \vec{y}$  ($\prec_s$ is a partial order!). In particular, let $c' \geq 0$ be the minimal value such that $\vec{x}'\prec_s c' \vec{y}$. Then $\vec{x}\prec_s \vec{x}' \Rightarrow \vec{x}\prec_s c' \vec{y}$, so that the minimal $c\geq 0$ such that $\vec{x}\prec_s c \vec{y}$ always satisfies $c\leq c'$. Now, recall that by definition $\pi/4 \geq \lambda^0_1 \geq 0$, and notice that if some component $n_j$ of $\vec{n}$ fulfills $|n_j| > 1$, then the maximal component of the reordered version \cite{smaj} of $\vec{\lambda}^0 + \pi/2\vec{n}$ is at least $3\pi/4$. Thus, because of (T1.$i$), $\vec{\lambda}^0 \prec_s \vec{\lambda}^0 + \pi/2\vec{n}$. Then (T1.$ii$) implies that $c_{(0,0,0)} \leq c_{\vec{n}}$. Therefore we can restrict our attention to vectors $\vec{n}$ with $|n_j|\leq 1$. A case by case check shows that the pre-costs $c_{\vec{n}}$ with $\vec{n} \in \{ (-1,-1,-1), (0,-1,0), (0,0,-1), (0,0,1)\}$ fulfill $c_{(-1,0,0)} \leq c_{\vec{n}}$, since (cf. point (T1.$ii$) above) $\vec{\lambda}^0 + \pi/2(-1,0,0) \prec_s \vec{\lambda}^0 + \pi/2\vec{n}$ \cite{example}. Similarly, we obtain that for the remaining vectors $\vec{n}$ with $|n_j|\leq 1$ the pre-costs satisfy $c_{(0,0,0)} \leq c_{\vec{n}}$, because $\vec{\lambda}^0 + \pi/2(0,0,0) \prec_s \vec{\lambda}^0 + \pi/2\vec{n}$. The only remaining configurations, with vectors $\vec{n}\in \{(-1,0,0), (0,0,0)\}$, are incomparable according to the $\prec_s$ relation ---unless $\lambda_1^0 + |\lambda_3^0| \leq \pi/4$, in which case we always obtain $c_{(0,0,0)}\leq c_{(-1,0,0)}$---, and this is why the optimization of Eq. (\ref{theorem1}) has to be performed. $\Box$

\vspace{2mm}

{\bf Corollary:} ($a$) When U is such that $\lambda_1^0+|\lambda_3^0|\leq \pi/4$, then the interaction cost is always given by 
\be
C_H(U)=c_{(0,0,0)}.
\ee
($b$) If, instead, $\lambda_1^0+|\lambda_3^0| < \pi/4$, then Hamiltonians $H$ and $H'$ always exist such that $C_H(U)=c^{\vec{h}}_{(0,0,0)} < c^{\vec{h}}_{(-1,0,0)}$ and $C_{H'}(U)=c^{\vec{h}'}_{(-1,0,0)} < c^{\vec{h}'}_{(0,0,0)}$.
 
\vspace{2mm}

{\em Proof:} ($a$) follows from the fact that $\lambda_1^0+|\lambda_3^0|\leq \pi/4 \Rightarrow \vec{\lambda}^0 \prec_s \vec{\lambda}^0 + \pi/2 (-1,0,0)$. (b) can be checked by considering $H$ and $H'$ given by $\vec{h}=\vec{\lambda}^0$ and $\vec{h}' = (\pi/2-\lambda_1^0, \lambda_2^0,-\lambda_3^0)$. $\Box$

\vspace{2mm}

In order to analyze Eq. (\ref{theorem1}) we first consider some examples. For the Ising interaction $H=h \sigma_3\otimes\sigma_3$ (equivalently $h \sigma_1\otimes\sigma_1$) and an arbitrary gate $U$, Eq. (\ref{theorem1}) reads (cf. Theorem 2 of \cite{Khaneja}),
\be
C_{h\sigma_1\otimes\sigma_1} (U) = \frac{\lambda_1^0 + \lambda_2^0 + |\lambda_3^0|}{h}.
\label{xx}
\ee
Let us now instead focus on three specific gates and arbitrary interactions. By $\ket{m}\otimes\ket{n}$ ($m,n=0,1$) we denote the computational basis of two-qubits. The CNOT gate is defined as
\be
\ket{m}\otimes\ket{n} \longrightarrow \ket{m}\otimes\ket{n\oplus m},
\ee
where $\oplus$ is sum modulo 2. Using the method described in Ref. \cite{Kraus} we obtain its canonical vector, $\vec{\lambda}^0=(\pi/4,0,0)$. Similarly, the SWAP gate,
\be
\ket{m}\otimes\ket{n} \longrightarrow \ket{n}\otimes\ket{m},
\ee
has vector $\vec{\lambda}^0 = (\pi/4,\pi/4,\pi/4)$. We also consider a third, intermediate gate $U_{XY}$ with $\vec{\lambda}^0 = (\pi/4,\pi/4,0)$, that corresponds to
\be
\ket{m}\otimes\ket{n} \longrightarrow i^{|m-n|}\ket{n}\otimes\ket{m}.
\ee
For these three gates we find
\bea
\C_{H}(\mbox{CNOT}) &=& \frac{\pi}{4}\frac{1}{h_1}, \label{cnot}\\
\C_{H}(U_{XY}) &=& \frac{\pi}{4}\frac{2}{h_1+h_2-|h_3|}, \label{utilde}\\
\C_{H}(\mbox{SWAP}) &=& \frac{\pi}{4}\frac{3}{h_1+h_2+|h_3|}. 
\label{swap}
\eea
With these examples at hand we make the following two observations. First, to any gate $U$ there corresponds a {\em natural interaction} $H_U$, with vector either $\vec{h}_1$ or $\vec{h}_2$ as defined in Theorem 1. This natural interaction allows to perform gate $U$ optimally without need to intermediately simulate another Hamiltonian and therefore the time inefficiencies inherent in the process of simulation are avoided. In this sense the natural interactions for the CNOT gate, gate $U_{XY}$ and the SWAP gate are, respectively, the Ising interaction $\sigma_1 \otimes \sigma_1$, the XY--model interaction $\sigma_1\otimes \sigma_1 +\sigma_2\otimes \sigma_2$ and the Heisenberg or exchange interaction $\sigma_1\otimes \sigma_1 +\sigma_2\otimes \sigma_2 + \sigma_3\otimes \sigma_3$.

The second observation is that for any fixed interaction $H$, e.g. $H=h\sigma_1\otimes\sigma_1$ as in Eq. (\ref{xx}), the interaction cost induces an order in the set of gates. For instance, according to Eq. (\ref{xx}), a SWAP is the most time--consuming gate when the Ising interaction is available. Eqs. (\ref{cnot}-\ref{swap}) also show, however, that such an order depends on the available interaction. Using the exchange interaction $H=\sigma_1\otimes\sigma_1+\sigma_2\otimes\sigma_2+\sigma_3\otimes\sigma_3$, $U_{XY}$ is twice as time-consuming as a SWAP gate. 

Let us move to Definition 2. It endows the set of non-local gates with a partial order structure based on the notion of interaction cost, but which is independent of any particular interaction. By comparing the resources required to perform two gates, such a partial order captures the intuition that some gates are more non-local than others.

We have already argued that no gate more non-local (i.e., more time-consuming for all interactions) than all the others exists. 
It is also easy to see that a gate $\alpha \vec{\lambda}^0$ is always less non-local than $\vec{\lambda}^0$ for any $\alpha \in [0,1]$, since the pre-costs are linear in $\alpha$. Next we present an analytical characterization of the partial order relation $V \leq U$ in a region of the set of two-qubit gates \cite{region}.

\vspace{2mm}

{\bf Theorem 2:} Let $U$ and $V$ be two two-qubit gates with corresponding ordered vectors $\vec{\lambda}_U^0$ and $\vec{\lambda}_V^0$ such that in both cases the restriction $\lambda_1^0+|\lambda_3^0| \leq \pi/4$ holds. Then gate $U$ is more non-local than gate $V$ if and only if $\vec{\lambda}_V^0 \prec_s \vec{\lambda}_U^0$,
\be
V \leq U ~~ \Leftrightarrow ~~ \vec{\lambda}_V^0 \prec_s \vec{\lambda}_U^0.
\ee

\vspace{2mm}

{\em Proof:} Recall that the restrictions on $\vec{\lambda}_U^0$ and $\vec{\lambda}_V^0$ imply, because of Corollary 1, that the interaction costs $\C_H(U)$ and $\C_H(V)$ are given, respectively, by the smallest $c_U, c_V \geq 0$ such that 
\bea
\vec{\lambda}_U^0 &\prec_s& c_U  \vec{h}, \label{U}\\
\vec{\lambda}_V^0 &\prec_s& c_V  \vec{h}.
\eea
Suppose first that $V\leq U$, that is, that for any Hamiltonian $H$ we have $C_H(V) \leq C_H(U)$. Then we also have $\vec{\lambda}_V^0 \prec_s \C_H(U)\vec{h}$. In particular, if we choose the interaction $H$ to have vector $\vec{h}=\vec{\lambda}_U^0$, we have $\C_{H}(V)\leq\C_H(U) = 1$ and $\vec{\lambda}_V^0 \prec_s \C_H(U) \vec{h} = \vec{\lambda}_U^0$, which proves the direct implication. The inverse implication follows from (T1.$ii$) of the proof of Theorem 1, which shows $\vec{\lambda}_V^0 \prec_s \vec{\lambda}_U^0$ implies that $\C_H(V) \leq \C_H(U)$ for all $H$. $\Box$

As an example of this result, we see that the $U_{XY}$ gate is more non-local  than the CNOT gate, and that, as it was to be expected, gates with sufficiently small components $|\lambda_i^0|$ are less non-local than those with large $|\lambda_i^0|$ \cite{analog}.

Finally, recall that entanglement can be used as a catalyzer for Hamiltonian simulation \cite{cata}. In particular, if to each of the two interacting qubits $A$ and $B$ we attach an extra qubit $A'$ and $B'$, where the pair $A'B'$ is in a maximally entangled state, Hamiltonian $H$ between $A$ and $B$ can be used to perform, without consuming the entanglement of $A'B'$, more powerful simulations than before, provided that fast LU are allowed in $AA'$ and $BB'$. Consequently, the interaction cost of gates is modified when not only LU, but also entangled ancillas are available.

In this work we have characterized the time-optimal synthesis of two-qubit unitary transformations using an arbitrary two-qubit Hamiltonian. In particular, the interaction cost $\C_H(H)$ has been computed and optimal protocols have been described. We have also characterized, in a region of the space of two-qubit gates, a partial order structure related to their degree of non-locality. These results can be applied to the study of the interaction cost for particular processes, such as the creation of a maximally entangled state \cite{Duer,Kraus} or the transmition of a classical or quantum bit of information from one qubit to another \cite{Hammerer}. All these discussions involve only two interacting qubits. It would be desirable to obtain a generalization to higher--dimensional systems. The lack of an analog to decomposition (\ref{deco}) in these cases is a serious drawback. Another interesting generalization consists in considering the asymptotic scenario, where the aim is to perform a large number of copies of the same gate.

This work was supported by the European Community under project EQUIP (contract IST-1999-11053), the ESF and the Institute for Quantum Information GmbH. G.V. is supported by grant HPMF-CT-1999-00200 (Marie Curie fellowship) of the European Community.


\begin{references}
\bibitem{smaj} The s-majorization relation can be extended to non-ordered vectors as follows. Given a vector $\vec{u}=(u_1,u_2,u_3)$, we construct a new vector $\vec{u}^s=(u_1^s,u_2^s,u_3^s)$, $u_1^s\geq u_2^s \geq |u_3^s|$ by first decreasingly reordering the modulus of the components $u_i$, and by then giving $u_3^s$ the sign of the product $u_1u_2u_3$.
Then, for any pair of vectors $\vec{u}$ and $\vec{v}$, $\vec{u} \prec_s \vec{v}$ denotes the set of inequalities (\ref{smaj}) applied to $\vec{u}^s$ and $\vec{v}^s$.
\bibitem{upto} By permuting the components of $\vec{\lambda}^0 + \pi/2\vec{n}$, as well as multiplying by $-1$ two of the components, other vectors can be build that also correspond to a decomposition of $U$. These are, nevertheless, equivalent to the original decomposition from the perspective of the s-majorization relation $\prec_s$, and need not be analyzed for the purposes of optimal gate synthesis.
\bibitem{clar}  $(\frac{\pi}{2}- \lambda_1^0,\lambda_2^0,-\lambda_3^0)$ is the reordered version \cite{smaj} of $\vec{\lambda}^0+\pi/2(-1,0,0)$.
\bibitem{example} Let us consider, as an example, $\vec{n}=(0,0,1)$. If we properly reorder \cite{smaj} the coefficients, we obtain the vector $(\pi/2-\lambda_1^0, \lambda_2^0, -\lambda_3^0)$ from $\vec{\lambda}^0 + \pi/2(-1,0,0)$ and the vector $(\pi/2+\lambda_3^0,\lambda_1^0,\lambda_2^0)$ from $\vec{\lambda}^0 + \pi/2(0,0,1)$. Then \bea
\pi/2-\lambda_1^0 &\leq& \pi/2+\lambda_3^0\nonumber \\
\pi/2-\lambda_1^0 + \lambda_2^0 +\lambda_3^0 &\leq& \pi/2+\lambda_3^0+\lambda_1^0-\lambda_2^0\nonumber \\
\pi/2-\lambda_1^0 + \lambda_2^0 -\lambda_3^0 &\leq& \pi/2+\lambda_3^0+\lambda_1^0+\lambda_2^0\nonumber, 
\eea
so that indeed $\vec{\lambda}^0 + \pi/2(-1,0,0)\prec_s \vec{\lambda}^0 + \pi/2(0,0,1)$ and therefore the pre-costs satisfy $c_{(-1,0,0)} \leq c_{(0,0,1)}$ (cf. point (T1.$ii$) in the proof of Theorem 1).
\bibitem{region} Contrary to the problem of Hamiltonian simulation, where the analogous partial order has been solved completely \cite{Bennett}, here we do not have a linear space of Hermitian operators (Hamiltonians), but the rather involved manifold $SU(4)$. In particular, in the region surrounding $\vec{\lambda}^0=(\pi/4,\pi/4,\pi/4)$ geometrical aspects of this manifold make the characterization very difficult. We have already met a manifestation of this fact in Eq. (\ref{theorem1}), where two pre-costs, instead of just one, have to be considered due to periodicity properties of unitary evolutions. The analytical characterization presented here is possible precisely away from $\vec{\lambda}^0=(\pi/4,\pi/4,\pi/4)$.
\bibitem{analog} Remarkably enough, the necessary and sufficient conditions for this partial order are completely analogous to those for efficient Hamiltonian simulation \cite{Bennett}. Such  a correspondence was, in principle, only to be expected for infinitesimal gates. It also holds for finite gates due to some convexity properties explored in Theorems 9 and 10 of \cite{Khaneja}.
\end{references}
\end{document}